\documentclass[11pt]{article}
\usepackage{graphicx}
\usepackage{epsfig}

\newcommand{\BABARPubYear}    {02}

\newcommand{\BABARProcNumber} {106}
\newcommand{\SLACPubNumber} {9567}

\input pubboard/babarsym

\long\def\inst#1{\par\nobreak\kern 4pt\nobreak
    {\it #1}\par\vskip 10pt plus 3pt minus 3pt}

\begin{document}
{\pagestyle{empty}

\begin{flushright}
SLAC-PUB-\SLACPubNumber \\
\babar-PROC-\BABARPubYear/\BABARProcNumber \\
November, 2002 \\
\end{flushright}

\par\vskip 4cm

\begin{center}
\Large \bf $B_d$ mixing measurements with the \babar\ detector
\end{center}
\bigskip

\begin{center}
\large 
C. Voena \\
Universit\`a ``La Sapienza'' and INFN  Roma\\
P.le Aldo Moro 2 00185 Rome, Italy\\
(for the \lbabar\ Collaboration)
\end{center}
\bigskip \bigskip

\title{$B_d$ mixing measurements with the \babar\ detector}

\begin{center}
\large \bf Abstract
\end{center}

The \BzBzb\ oscillation frequency $\Delta m_d$ has been measured with the \babar\ detector at the PEP-II asymmetric $B$ factory with
different experimental techniques. The discussion here is focused on the recent simultaneous measurement of $\Delta m_d$ and $\tau_{B^0}$ with exclusively reconstructed $B^0 \rightarrow D^{*-}\ell^+\nu_l$ decays, based on 23 million $B\bar{B}$ pairs collected by \babar\ .
The measurements of $\Delta m_d$ with fully reconstructed hadronic decays and
with dilepton events are also reviewed. The average \babar\ result is $\Delta m_d = 0.500 \pm 0.008 \pm 0.006$ $ps^{-1}$.

\vfill
\begin{center}
Contributed to the Proceedings of the 31$^{st}$ International 
Conference on High Energy Physics, \\
7/24/2002---7/31/2002, Amsterdam, the Netherlands
\end{center}

\vspace{1.0cm}
\begin{center}
{\em Stanford Linear Accelerator Center, Stanford University, 
Stanford, CA 94309} \\ \vspace{0.1cm}\hrule\vspace{0.1cm}
Work supported in part by Department of Energy contract DE-AC03-76SF00515.
\end{center}

\section{Introduction}
Particle-antiparticle oscillations or mixing,
has been observed in the neutral $B\bar{B}$ meson system
almost fifteen years ago~\cite{argus}.
This quantum-mechanical behavior is originated by the flavor eigenstates
$B^0$ and $\bar{B}^0$ not being Hamiltonian eigenstates. The frequency of
the oscillation is the mass difference between the mass eigenstates, $\Delta m_d$.
In the Standard Model, $B^0\bar{B}^0$ mixing occours through second-order  weak diagrams involving the exchange of up-type quarks, with the top quark contributing the dominant amplitude. A measurement of $\Delta m_d$
is therefore sensitive to the value of the Cabibbo-Kobayashi-Maskawa matrix element $V_{td}$~\cite{vtd}.
The oscillation frequency $\Delta m_d$ has been measured with both time-integrated and time-dependent techniques ~\cite{dmold}.
Asymmetric $B$ factory experiments like \babar\ can perform high statistics time dependent measurements of $\Delta m_d$.
\section{Measurements of $\Delta m_d$ with the \babar\ detector}
The \babar\ detector~\cite{detector}  collects data at the PEP-II
asymmetric $e^+e^-$ collider operated at or near the $\Upsilon(4S)$ resonance.
$B\bar{B}$ pairs from $\Upsilon(4S)$ decay move along the high-energy beam
direction ($z$) with a nominal Lorentz boost $<\beta\gamma> = 0.55$. Therefore,
the two $B$ decays vertices are separated by about 260 $\mu m$ on average.
The two $B$ mesons are produced in a coherent $P$-wave state and their
proper decay-time difference $\Delta t$ distribution is governed by the following probabilities to observe $mixed$(-) or $unmixed$(+) events:\\
\begin{eqnarray}
&&Prob(B^0\bar{B}^0\rightarrow B^0\bar{B}^0, B^0B^0 or \bar{B}^0\bar{B}^0)\propto \nonumber \\ && e^{-\frac{|\Delta t|}{\tau_{B^0}}}(1\pm cos\Delta m_d \Delta t).
\label{eq:alltrue}
\end{eqnarray}
Therefore, a measurement of $\Delta t$ together with the identification
of the $b$-flavor of both $B$ mesons at their time of decay, allows to
observe the oscillations and to extract $\Delta m_d$.

In the following sections the simultaneous
 measurement of $\Delta m_d$ and $\tau_{B^0}$ with exclusively reconstructed $B^0 \rightarrow D^{*-}\ell^+\nu_l$ decays~\cite{dslnu}, the measurement of $\Delta m_d$ with fully reconstructed hadronic $B^0$ decays~\cite{hadronic} and with inclusively reconstructed dilepton events~\cite{dilepton}, will be discussed.
\section{Measurement of $\Delta m_d$ and $\tau_{B^0}$ with exclusively reconstructed $B^0 \rightarrow D^{*-}\ell^+\nu_l$ decays}\label{dst}
The analysis is based on a sample of approximately 14,000 exclusively
reconstructed $B^0 \rightarrow D^{*-}\ell^+\nu_l$ decays selected from 23 million $B\bar{B}$ pairs recorded in the years 1999-2000 by \babar\ .
The purity of the sample is 65-89$\%$ depending on the decay mode of the $\bar{D}^0$ from the $D^{*-}$.

One of the two $B$ produced by the $\Upsilon(4S)$ decay is reconstructed in the
semileptonic mode and the charge of the final-state particles identifies the
flavor of the $B$.
$D^{*-}$ candidates are reconstructed using the decay $D^{*-} \rightarrow \bar{D}^0 \pi^{-}$, while $\bar{D}^0$ candidates are reconstructed in the modes  $K^{+}\pi^{-}$, $K^{+}\pi^{-}\pi^{0}$, $K^{+}\pi^{-}\pi^{+}\pi^{-}$ and $K^0_s\pi^{+}\pi^{-}$.
$D^{*-}$ candidates are then combined with oppositely charged high-energy electrons or muons in the event and the $D^{*-}\ell^+$ pair is required to pass kinematic cuts that enhance the contribution of $B^0 \rightarrow D^{*-}\ell^+\nu_l$ decays.
The $B$ mass and energy cannot be reconstructed  because of the presence of the
neutrino, thus the distribution $\delta m$ of the difference between the $D^{*-}$ and the $D^0$ masses is used to select $B$ candidates.
The distribution of $\delta m$ for events passing the selection criteria in the muon sample is shown in Figure ~\ref{fig:dm}.

\begin{figure}[!tbp]
\vspace{9pt}
\centerline {
\epsfysize=6.0cm
\epsfbox{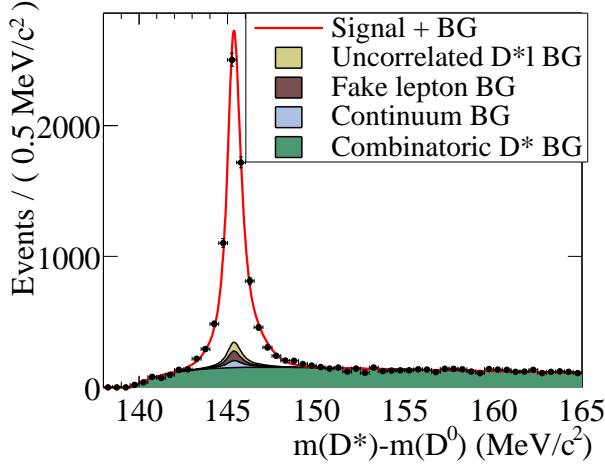}}
\caption{$\delta m$ distribution for events passing all selection criteria for $B^0 \rightarrow D^{*-}\ell^+\nu_l$ decays with a muon candidate. The points correspond to the data. The curve is the result of a fit. The shaded distributions correspond to the four types of background (BG) described in the text.}
\label{fig:dm}
\end{figure}

There are two types of background with respect to the $\delta m$ distribution: combinatoric background and peaking background.
Combinatoric background is due to events with a mis-reconstructed $D^{*-}$
and does not peak in the $\delta m$ distribution. Peaking background is due to $B\bar{B}$ events with mis-identified leptons or uncorrelated true leptons, and continuum events. The $\delta m$ distribution of the peaking background is the same as the distribution of the signal.
Several control samples are selected to characterize the various backgrounds in both the fraction and the time distribution.

All the charged tracks in the event, except the reconstructed tracks from the $D^{*-}\ell^+$  pair, are used to identify the flavor of the other $B$ (referred to as $B_{tag}$). There are five types of tagging categories. The first two tagging categories rely on the presence
of a prompt lepton or charged kaons in the event, whose charge is correlated with the $b$-flavor of the decaying B.
The other three categories exploit a variety of inputs (e.g. slow pions, momentum of the track with the maximum center-of-mass momentum) with a neural network
technique.

The difference $\Delta t$ between the two $B$ decay times is determined from the measured separation $\Delta z = z_{D*l}-z_{tag}$ along the beam axis between the $D^{*-}\ell^+$ vertex and the $B_{tag}$ vertex.
The measured $\Delta z$ is converted into $\Delta t$ with the known $\Upsilon(4S)$ boost according to the relation $\Delta z = c \beta \gamma \Delta t$ which neglects the small $B$ momentum in the $\Upsilon(4S)$ frame. The resolution on the
$D^{*-}\ell^+$ vertex is about 70 $\mu m$ while the resolution on the $B_{tag}$
vertex is about 160 $\mu m$.

Each tagging category $i$ has a probability $w_i$ of incorrectly assigning the flavor of the $B_{tag}$ and there is a limited precision on the $\Delta t $ measurement. These two experimental complications affect the $\Delta t $ distribution of
Equation ~\ref{eq:alltrue} which becomes:

\begin{eqnarray}
&&Prob(B^0\bar{B}^0 \rightarrow B^0\bar{B}^0, B^0B^0 or \bar{B}^0\bar{B}^0)
\propto \nonumber  \\&& {\mathcal R}(\delta t;{\hat a}) \otimes e^{-\frac{|\Delta t|}{\tau_{B0}}} (1\pm (1-2w_i)cos\Delta m_d \Delta t). \nonumber
\label{ }
\end{eqnarray}

The function ${\mathcal R}(\delta t;{\hat a})$ is the resolution function
which parametrizes the response to $\Delta t$ of the detector, $\delta t$ =
$\Delta t_{meas}$ - $\Delta t_{true}$ and ${\hat a }$ is a set of parameters.
The final $\Delta t$ distribution also includes terms for each relevant background source.

The oscillation frequency $\Delta m_d$ and the lifetime $\tau_{B^0}$
are determined simultaneously with an unbinned maximum likelihood fit to the
measured $\Delta t $ distribution. Note that other mixing measurements fix $\tau_{B^0}$ to the world average and this is the source of the dominant systematic error. Also the resolution, the fraction of charged $B$, the mistag and the background parameters are floated in the fit. The results are: $\Delta m_d$ = $0.492 \pm 0.018(stat) \pm 0.013(syst)$ $ps^{-1}$ and $\tau_{B^0}$ = $1.523^{+0.024}_{-0.023}(stat) \pm 0.022(syst)$ $ps$. The correlation between $\Delta m_d$ and $\tau_{B^0}$ is -0.22. A correction is applied to both $\Delta m_d$ and $\tau_{B^0}$ which takes into account selection and fit biases. The uncertainty on such a correction is the
dominant systematic error.
Other \babar\ measurements for $\tau_{B^0}$ can be found in ~\cite{taupart} and
~\cite{tauhad}.
Figure ~\ref{fig:asym} shows the mixing asymmetry defined as the difference between the number of unmixed and mixed events over their sum as a function of $\Delta t$.

\begin{figure}[!tbp]
\vspace{9pt}
\centerline {
\epsfysize=6.0cm
\epsfbox{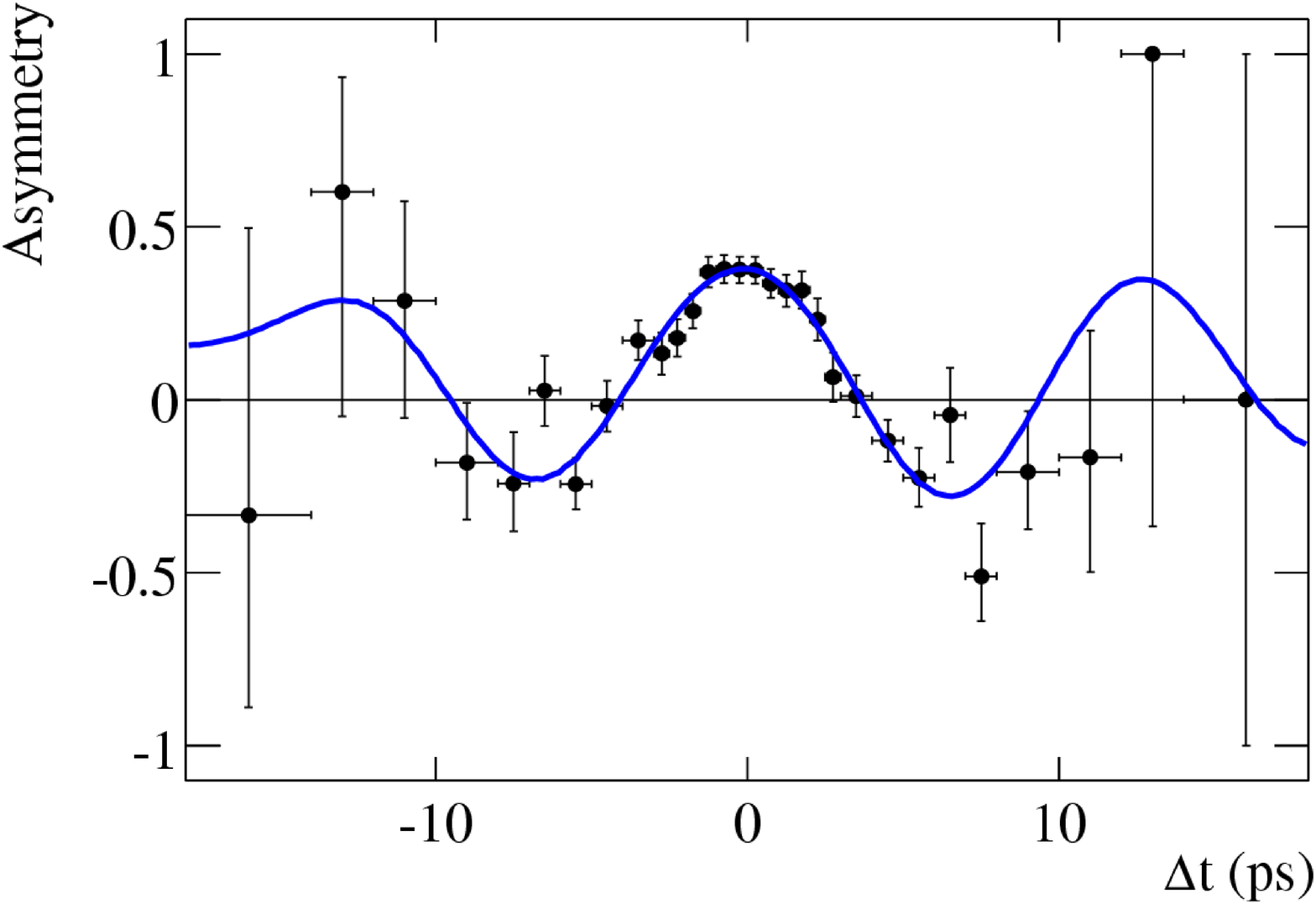}}
\caption{Mixing asymmetry plot for a 80$\%$ pure $B^0 \rightarrow D^{*-}\ell^+\nu_l$ sample. The dots are the data and the curve is the projection of the fit result.}
\label{fig:asym}
\end{figure}

\section{Measurement of $\Delta m_d$ with fully reconstructed $B^0$ hadronic decays}
The analysis is based on $\sim$6300 $B^0$ selected from 32 million $B\bar{B}$ pairs. The $B^0$ are reconstructed in the flavor eigenstates $D^{*-}\pi^+$, $D^{*-}\rho^+$, $D^{*-}a_1^+$, $J/\psi K^{*0}$ and the purity of
the selected sample is about 86$\%$.
$B^0$ candidates are selected using the difference between the energy of the candidate and the beam energy $\sqrt{s}$/2 in the center of mass frame, and the beam energy substituted mass, calculated from $\sqrt{s}$/2 and the reconstructed momentum of the $B$. 

The tagging algorithm and the tagging vertex reconstruction technique are described in the previous section.
$\Delta m_d$ is extracted with an unbinned maximum likelihood fit to the
$\Delta t$ distribution (obtained from $\Delta z$ as described in the previous section) where also all the mistag probabilities and the resolution parameters are floated.
The result is $\Delta m_d = 0.516 \pm 0.016(stat) \pm 0.010(syst)$ $ps^{-1}$, the dominant systematic error being the uncertainty on $\tau_{B^0}$ which is fixed in the fit.

\section{Measurement of $\Delta m_d $ with inclusive dilepton  events}
The analysis is based on 23 million $B\bar{B}$ pairs.
The measurement technique consists in the identification of events containing
two high energy leptons from semileptonic decays of $B$ mesons. The flavor of the $B$ mesons at the time of their decay is determined by the charge of the
leptons.
About 99000 events are selected, $\sim$ 55 $\%$ of them being $B^+B^-$ events
which are not removed by the event selection criteria. Another non negligible
background is due to leptons from the $b\rightarrow c\rightarrow l$ decay chain
($\sim$ 13 $\%$) which are also the main source of wrongly tagged events. The difference of the $z$ coordinates of the two $B$ decay vertices $\Delta z$ is determined using the two lepton tracks
and a beam spot constraint. $\Delta t$ is then obtained from $\Delta z$ as described in section ~\ref{dst}.
A binned maximum likelihood fit is used to extract $\Delta m_d$ together
with the resolution parameters, the fraction of charged $B$ and some of the
background fractions and parameters. The result is $\Delta m_d = 0.493 \pm 0.012(stat) \pm 0.009(syst)$ $ps^{-1}$. The dominant systematic error is due to the uncertainty on $B^0$ and $B^+$ lifetimes which are fixed in the fit.

\section{Conclusions}
The oscillation frequency $\Delta m_d$ of the \BzBzb\ system has been
measured by the \babar\ experiment with different experimental techniques.
The combined \babar\ result is
$\Delta m_d = 0.500 \pm 0.008 \pm 0.006$ $ps^{-1}$. The corresponding precision is $2\%$ to be compared with the world average precision which is 1.2 $\%$.

\end{document}